\documentclass[aps,pre,
showkeys,
longbibliography,
reprint,                
twocolumn,              
nofootinbib,
floatfix]{revtex4-2}
\usepackage[T1]{fontenc}
\usepackage[utf8]{inputenc}
\usepackage{amsmath}
\usepackage{amsfonts}
\usepackage{graphicx}
\usepackage{rotating}
\usepackage{xcolor}
\usepackage{subfig}
\usepackage[colorlinks=true,hyperfootnotes=true,breaklinks=true,allcolors=teal]{hyperref}
\usepackage[capitalise]{cleveref}

\begin{document}

\title{Random site percolation with complex neighborhoods in five dimension}

\author{Krzysztof Malarz}
\thanks{ORCID~\href{https://orcid.org/0000-0001-9980-0363}{0000-0001-9980-0363}}
\email{malarz@agh.edu.pl}
\affiliation{\href{https://ror.org/00bas1c41}{AGH University}, Faculty of Physics and Applied Computer Science, al.~Mickiewicza~30, 30-059 Krak\'ow, Poland}

\author{Maciej Wo{\l}oszyn}
\thanks{ORCID~\href{https://orcid.org/0000-0001-9896-1018}{0000-0001-9896-1018}}
\email{woloszyn@agh.edu.pl}
\affiliation{\href{https://ror.org/00bas1c41}{AGH University}, Faculty of Physics and Applied Computer Science, al.~Mickiewicza~30, 30-059 Krak\'ow, Poland}

\begin{abstract}
In this paper, the random site percolation problem in a five-dimensional space for complex neighborhoods is studied.
The efficient C++ code (with ordinary work division) of the classical Newman--Ziff algorithm is presented. 
The obtained speed-up of computations reduces 2170 years of single-core computations---necessary for obtaining the results presented in this paper---much below the typical half-decay time of the scientist.
For complex neighborhoods, it is for neighborhoods composed with sites taken from several coordination zones (up to the seventh coordination zone), the 127 percolation thresholds are calculated (with 120 among them being estimated for the first time).
For seven extended (compact) neighborhoods, the fractal dimensions are also calculated.
The mean value of these fractal dimensions, averaged over these seven compact neighborhoods, is estimated as $\langle d_f\rangle\approx 3.5581(70)$.
The percentage errors of the values obtained for the fractal dimensions for compact neighborhoods vary from 0.26\% to 1.76\% with respect to the theoretically predicted value based on scaling relations and the most recent estimates of critical exponents for five-dimensional space.
The universality of the percolation threshold as dependent on the weighted coordination number $\zeta=\sum_i z_i r_i$ (where $z_i$ is the number of sites in the $i$-th coordination zone and $r_i$ is the Euclidean distance from the sites in the $i$-th coordination zone to the central site) is also verified.
The latter manifests itself as the power law ($p_c\propto\zeta^{-g}$) with {$g\approx 0.7913(43)$}.
\end{abstract}

\date{July 21, 2026}

\maketitle

\section{Introduction}

The percolation \cite{bookDS,bookBB,bookMS,bookHK} is a core problem in statistical mechanics that offers a look inside the phase transition solely on geometrical consideration.
Originating from \citeauthor{Broadbent1957} \cite{Broadbent1957,Hammersley1957} works devoted to rheology (and still scientifically interesting and explored  \cite{RevModPhys.65.1393,Berkowitz_1993,ISI:000524118200031}), researchers quickly found plenty of obvious applications of percolation in the physics of 
ferromagnetic phase transition \cite{Fortuin_1972,Coniglio_1980,Bialas_2000}, 
understanding properties of magnetic materials \cite{PhysRevB.94.054407,Alguero_2020}
gelation and sol--gel transitions \cite{Stauffer_1976,PhysRevE.63.011510}, 
metal--insulator transitions \cite{Efros_1976,McLachlan_1987} 
and electrical conductivity \cite{RevModPhys.45.574,Tsangaris_2002}, 
but also in less obvious fields, including: 
construction and building materials \cite{Qing_2026};
forest fires \cite{PhysRevLett.69.1629,Clar_1996,Kaczanowska2002};
epidemic spreading \cite{PhysRevE.66.016128,Meyers_2007,2101.00550}; 
agriculture \cite{PhysRevE.101.032301,PhysRevE.109.014304,Alonso_Tlali_2025}; 
etc. (see References~\onlinecite{Saberi2015,Li_2021} for recent reviews).

In the geometrical formulation of percolation one deals with a system of occupied and empty sites (in site percolation problem) or open and closed bonds (in bond percolation problem).
The sites (bonds) are occupied (open) with probability $p$ and empty (close) with probability $1-p$.
Depending on the geometry of the system, there is a critical probability $p_c$, above which the occupied sites (open bonds) create a path connecting the borders of the system.
In other words, at $p=p_c$ the cluster of occupied sites (open bonds) that span the system appears for the first time.  
This cluster at $p=p_c$ is termed an incipient percolation cluster and has fractal properties \cite{Stauffer_1979}.
The critical probability $p_c$ is called the percolation threshold and is one of the most important characteristics of percolating systems.

Usually, two occupied sites (open bonds) are considered to belong to the same cluster when they are the nearest neighbors.
But since \citeauthor{Dalton_1964} \cite{Dalton_1964} and \citeauthor{Domb1966} \cite{Domb1966} works also long-range interactions (extensively studied by \citeauthor{Iribarne1999} \cite{Iribarne1999}) are considered, where clusters of occupied sites (open bonds) are connected even when their geometrical distance is larger than the lattice constant.
Such cases are concerned with percolation with extended (when the neighborhood of sites is compact) or complex (when the neighborhood of sites is non-compact, with holes) neighborhoods.
Later percolation thresholds were estimated for complex (or at least extended) neighborhoods in
two- (square \cite{Galam2005a,Galam2005b,Majewski2007,2010.02895,2303.10423,2310.20668,2503.16703}, 
triangular \cite{2006.15621,2102.10066,2310.20668} 
and honeycomb \cite{2204.12593,2310.20668}),
three- \cite{Kurzawski2012,Malarz2015,2010.02895},
four- \cite{1803.09504},
and five-dimensional \cite{PhysRevE.98.022120,2308.15719} lattices.

The other path of studies on percolation phenomenon is the search for a universal formula for $p_c$. 
The search for the universal formula for $p_c$ based solely on the dimension $d$ in which the percolating system is embedded and the topology of the network has a long tradition, pointing out the works of \citeauthor{Galam_Mauger_1994b} \cite{Galam_Mauger_1994b,PhysRevE.53.2177,Galam_1997322,Galam_1998255}. 
In these searches, the lattice topology is usually described by its dimension and the lattice connectivity, {\it i.e.}, the number of neighbors in the assumed neighborhood.
For example, \citeauthor{PhysRevE.105.024105} \cite{PhysRevE.105.024105} found that for the site percolation problem, the percolation threshold $p_c$ follows asymptotically
\begin{equation}
\label{eq:pc-discs}
p_c\propto \frac{1}{z}.
\end{equation}
Adding the term $b$ to the denominator of \Cref{eq:pc-discs}
\begin{equation}
\label{eq:pc_const-over-z}
p_c = \frac{c}{z+b}
\end{equation}
allows for taking into account the finite-$z$ effect \cite{2010.02895}. 
The third universal scaling studied in Reference~\onlinecite{PhysRevE.105.024105} was
\begin{equation}
\label{eq:pc-exp}
p_c= 1-\exp(d/z)
\end{equation}
proposed by \citeauthor{Koza_2014} \cite{Koza_2014,Koza_2016}.
Some heuristic dependencies on the weighted coordination number $\zeta$ such as
\begin{equation} 
\label{eq:pc_zeta-to-gamma}
p_c\propto\zeta^{-g}
\end{equation}
were also proposed and studied \cite{2310.20668}.
For two-dimensional lattices $g\approx 1/2$ was observed \cite{2204.12593}.
We also note other attempts \cite{Koza_2014,Koza_2016,PhysRevE.103.022126,PhysRevE.103.022127,PhysRevE.105.024105} to search universal formulas that allow prediction of $p_c$ based solely on the lattice characteristics.

In this paper, we go a step further and evaluate the percolation thresholds $p_c$ for complex neighborhoods for a five-dimensional simple cubic lattice with neighborhoods comprising combination of sites from the first and up to seventh coordination zones.
Some dependencies of the obtained percolation threshold on the lattice characteristics, namely, for weighted coordination number, are proposed.
Finally, the fractal dimension of the incipient percolation cluster is also evaluated for extended (compact) neighborhoods.

The paper is organized as follows: in \Cref{sec:Methods} methods for calculating the percolation thresholds, the source code, and its possible primitive but efficient parallelization are presented.
\Cref{sec:Results} shows the results of Monte Carlo simulations for calculating the percolation thresholds for complex neighborhoods and the fractal dimension of an incipient percolation cluster in the case of extended (compact) neighborhoods.
Some dependencies of percolation thresholds on lattice characteristics are also presented there.
\Cref{sec:Discussion} is devoted to a discussion of the results obtained.
The source of the computer program and dozens of detailed results (for both percolation thresholds and fractal dimensions) are presented in the Supplemental Material \cite{SM}.

\begin{figure}[tbp]
\includegraphics[width=.95\columnwidth]{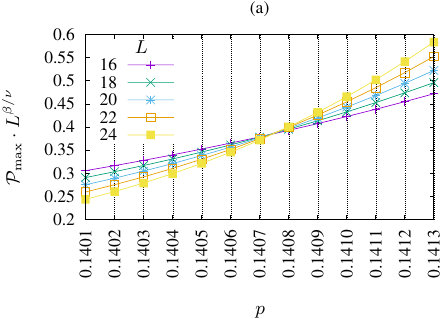}
\includegraphics[width=.95\columnwidth]{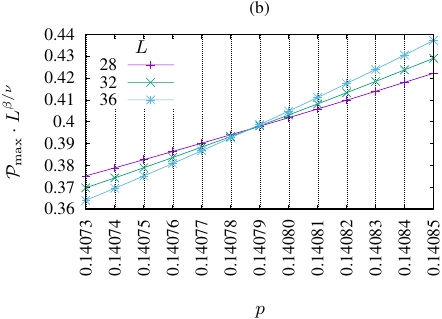}
\caption{\label{fig:example_sc5-rescaledPmax_vs_p}Example of rescaled probability of belonging to the largest cluster $\mathcal{P}_\text{max}\cdot L^{\beta/\nu}$ as dependent on the site occupation probability $p$ for \textsc{sc(5)-1} lattice (see 
Figure 5
in Supplemental Material \cite{SM} for remaining cases).
(a) $R=10^5$ repetitions of the system construction give clear common crossing point for various lattice sizes $L$ with $10^{-4}$ accuracy, (b) while ten times better statistics ($R=10^6$) improves this accuracy to $10^{-5}$}
\end{figure}

\begin{table}[!h]
\caption{\label{tab:pc}Percolation thresholds $p_c$ for various complex neighborhoods in simple cubic five-dimensional lattice}
\begin{ruledtabular}
\begin{tabular}{rrrl}
lattice \textsc{sc(5)-} & $z$ & $\zeta$ & $p_c$ \\
\hline 
           1 & $10$ & $10.0000$ & $0.14079$\footnote{$0.14079633(4)$ \cite{PhysRevE.98.022120}}\\
\hline 
           2 & $40$ & $40\sqrt{2}\approx 56.5685$ & $0.04357$\\
         1,2 & $50$ & $66.5685$ & $0.03379$\footnote{$0.033794(4)$ \cite{2308.15719}}\\
\hline 
            3 &  $80$ & $80\sqrt3\approx 138.5641$ & $0.01663$\\
          1,3 &  $90$ & $148.5641$ & $0.01562$\\
          2,3 & $120$ & $195.1326$ & $0.01351$\\
        1,2,3 & $130$ & $205.1326$ & $0.01318$\footnote{$0.013190(2)$ \cite{2308.15719}}\\
\hline 
            4 &  $90$ & $180.0000$ & $0.01561$\\
          1,4 & $100$ & $190.0000$ & $0.01347$\\
          2,4 & $130$ & $236.5685$ & $0.01311$\\
          3,4 & $170$ & $318.5641$ & $0.00877$\\
        1,2,4 & $140$ & $246.5685$ & $0.01148$\\
        1,3,4 & $180$ & $328.5641$ & $0.00862$\\
        2,3,4 & $210$ & $375.1326$ & $0.00792$\\
      1,2,3,4 & $220$ & $385.1326$ & $0.00784$\footnote{$0.007850(1)$ \cite{2308.15719}}\\
\hline 
            5 & $112$ & $112\sqrt{5}\approx 250.4396$ & $0.01063$\\
          1,5 & $122$ & $260.4396$ & $0.01004$\\
          2,5 & $152$ & $307.0082$ & $0.00881$\\
          3,5 & $192$ & $389.0037$ & $0.00703$\\
          4,5 & $202$ & $430.4396$ & $0.00685$\\
        1,2,5 & $162$ & $317.0082$ & $0.00852$\\
        1,3,5 & $202$ & $399.0037$ & $0.00686$\\
        1,4,5 & $212$ & $440.4396$ & $0.00671$\\
        2,3,5 & $232$ & $445.5722$ & $0.00658$\\
        2,4,5 & $242$ & $487.0082$ & $0.00621$\\
        3,4,5 & $282$ & $569.0037$ & $0.00546$\\
      1,2,3,5 & $242$ & $455.5722$ & $0.00647$\\
      1,2,4,5 & $252$ & $497.0082$ & $0.00611$\\
      1,3,4,5 & $292$ & $579.0037$ & $0.00541$\\
      2,3,4,5 & $322$ & $625.5722$ & $0.00518$\\
    1,2,3,4,5 & $332$ & $635.5722$ & $0.00515$\footnote{$0.0051575(5)$ \cite{2308.15719}}\\
\hline 
            6 & $240$ & $240\sqrt{6}\approx 587.8775$ & $0.00582$\\
          1,6 & $250$ & $597.8775$ & $0.00538$\\
          2,6 & $280$ & $644.4461$ & $0.00548$\\
          3,6 & $320$ & $726.4416$ & $0.00436$\\
          4,6 & $330$ & $767.8775$ & $0.00482$\\
          5,6 & $352$ & $838.3172$ & $0.00381$\\
        1,2,6 & $290$ & $654.4461$ & $0.00508$\\
        1,3,6 & $330$ & $736.4416$ & $0.00429$\\
        1,4,6 & $340$ & $777.8775$ & $0.00448$\\
        1,5,6 & $362$ & $848.3172$ & $0.00378$\\
        2,3,6 & $360$ & $783.0101$ & $0.00415$\\
\end{tabular}
\end{ruledtabular}
\end{table}
\begin{table}\ContinuedFloat
\caption{continued}
\begin{ruledtabular}
\begin{tabular}{rrrl}
lattice \textsc{sc(5)-} & $z$ & $\zeta$ & $p_c$ \\ \hline
        2,4,6 & $370$ & $824.4461$ & $0.00467$\\
        2,5,6 & $392$ &  $894.8857$ & $0.00365$\\
        3,4,6 & $410$ &  $906.4416$ & $0.00373$\\
        3,5,6 & $432$ &  $976.8812$ & $0.00343$\\
        4,5,6 & $442$ & $1018.3172$ & $0.00333$\\
      1,2,3,6 & $370$ &  $793.0101$ & $0.00411$\\
      1,2,4,6 & $380$ &  $834.4461$ & $0.00434$\\
      1,2,5,6 & $402$ &  $904.8857$ & $0.00362$\\
      1,3,4,6 & $420$ &  $916.4416$ & $0.00369$\\
      1,3,5,6 & $442$ &  $986.8812$ & $0.00340$\\
      1,4,5,6 & $452$ & $1028.3172$ & $0.00331$\\
      2,3,4,6 & $450$ &  $963.0101$ & $0.00362$\\
      2,3,5,6 & $472$ & $1033.4498$ & $0.00332$\\
      2,4,5,6 & $482$ & $1074.8857$ & $0.00324$\\
      3,4,5,6 & $522$ & $1156.8812$ & $0.00305$\\
    1,2,3,4,6 & $460$ &  $973.0101$ & $0.00359$\\
    1,2,3,5,6 & $482$ & $1043.4498$ & $0.00330$\\
    1,2,4,5,6 & $492$ & $1084.8857$	& $0.00322$\\ 
    1,3,4,5,6 & $532$ & $1166.8812$	& $0.00304$\\
    2,3,4,5,6 & $562$ & $1213.4498$ & $0.00298$\\
  1,2,3,4,5,6 & $572$ & $1223.4498$ & $0.00298$\footnote{$0.002981(1)$ \cite{2308.15719}}\\
\hline 
            7 & $320$ & $320\sqrt7\approx 846.6404$ & $0.00378$\\
          1,7 & $330$ &  $856.6404$ & $0.00371$\\
          2,7 & $360$ &  $903.2090$ & $0.00358$\\
          3,7 & $400$ &  $985.2045$ & $0.00325$\\
          4,7 & $410$ & $1026.6404$ & $0.00325$\\
          5,7 & $432$ & $1097.0800$ & $0.00298$\\
          6,7 & $560$ & $1434.5180$ & $0.00248$\\
        1,2,7 & $370$ &  $913.2090$ & $0.00352$\\
        1,3,7 & $410$ &  $995.2045$ & $0.00321$\\
        1,4,7 & $420$ & $1036.6404$ & $0.00321$\\
        1,5,7 & $442$ & $1107.0800$ & $0.00295$\\
        1,6,7 & $570$ & $1444.5180$ & $0.00246$\\
        2,3,7 & $440$ & $1041.7730$ & $0.00315$\\
        2,4,7 & $450$ & $1083.2090$ & $0.00312$\\
        2,5,7 & $472$ & $1153.6486$ & $0.00291$\\
        2,6,7 & $600$ & $1491.0865$ & $0.00241$\\  
        3,4,7 & $490$ & $1165.2045$ & $0.00293$\\
        3,5,7 & $512$ & $1235.6441$ & $0.00268$\\
        3,6,7 & $640$ & $1573.0820$ & $0.00231$\\
        4,5,7 & $522$ & $1277.0800$ & $0.00270$\\
        4,6,7 & $650$ & $1614.5180$ & $0.00227$\\
        5,6,7 & $672$ & $1684.9576$ & $0.00218$\\
      1,2,3,7 & $450$ & $1051.7730$ & $0.00311$\\
      1,2,4,7 & $460$ & $1093.2090$ & $0.00308$\\
      1,2,5,7 & $482$ & $1163.6486$ & $0.00287$\\
\end{tabular}
\end{ruledtabular}
\end{table}
\begin{table}\ContinuedFloat
\caption{continued}
\begin{ruledtabular}
\begin{tabular}{rrrl}
lattice \textsc{sc(5)-} & $z$ & $\zeta$ & $p_c$ \\ \hline
\
      1,2,6,7 & $610$ & $1501.0865$ & $0.00239$\\
      1,3,4,7 & $500$ & $1175.2045$ & $0.00290$\\
      1,3,5,7 & $522$ & $1245.6441$ & $0.00266$\\
      1,3,6,7 & $650$ & $1583.0820$ & $0.00230$\\
      1,4,5,7 & $532$ & $1287.0800$ & $0.00268$\\
      1,4,6,7 & $660$ & $1624.5180$ & $0.00226$\\
      1,5,6,7 & $682$ & $1694.9576$ & $0.00218$\\
      2,3,4,7 & $530$ & $1221.7730$ & $0.00284$\\
      2,3,5,7 & $552$ & $1292.2126$ & $0.00264$\\ 
      2,3,6,7 & $680$ & $1629.6506$ & $0.00226$\\
      2,4,5,7 & $562$ & $1333.6486$ & $0.00264$\\
      2,4,6,7 & $690$ & $1671.0865$ & $0.00222$\\
      2,5,6,7 & $712$ & $1741.5261$ & $0.00214$\\
      3,4,5,7 & $602$ & $1415.6441$ & $0.00249$\\
      3,4,6,7 & $730$ & $1753.0820$ & $0.00215$\\
      3,5,6,7 & $752$ & $1823.5216$ & $0.00207$\\
      4,5,6,7 & $762$ & $1864.9576$ & $0.00204$\\
    1,2,3,4,7 & $540$ & $1231.7730$	& $0.00283$\\
    1,2,3,5,7 & $562$ & $1302.2126$	& $0.00262$\\
    1,2,3,6,7 & $690$ & $1639.6506$ & $0.00225$\\
    1,2,4,5,7 & $572$ & $1343.6486$	& $0.00262$\\
    1,2,4,6,7 & $700$ & $1681.0865$ & $0.00221$\\
    1,2,5,6,7 & $722$ & $1751.5261$ & $0.00214$\\
    1,3,4,5,7 & $740$ & $1763.0820$ & $0.00214$\\
    1,3,4,6,7 & $612$ & $1425.6441$ & $0.00248$\\
    1,3,5,6,7 & $762$ & $1833.5216$ & $0.00206$\\
    1,4,5,6,7 & $772$ & $1874.9576$ & $0.00203$\\
    2,3,4,5,7 & $642$ & $1472.2126$ & $0.00246$\\
    2,3,4,6,7 & $770$ & $1809.6506$ & $0.00211$\\
    2,3,5,6,7 & $792$ & $1880.0902$ & $0.00204$\\
    2,4,5,6,7 & $802$ & $1921.5261$	& $0.00201$\\
    3,4,5,6,7 & $842$ & $2003.5216$ & $0.00195$\\
  1,2,3,4,5,7 & $652$	& $1482.2126$ & $0.00245$\\
  1,2,3,4,6,7 & $780$	& $1819.6506$ & $0.00211$\\
  1,2,3,5,6,7 & $802$	& $1890.0902$ & $0.00204$\\
  1,2,4,5,6,7 & $812$ & $1931.5261$ & $0.00200$\\
  1,3,4,5,6,7 & $852$ & $2013.5216$ & $0.00195$\\
  2,3,4,5,6,7 & $882$ & $2060.0902$ & $0.00193$\\
1,2,3,4,5,6,7 & $892$ & $2070.0902$ & $0.00193$\footnote{$0.001934(1)$ \cite{2308.15719}}\\
\end{tabular}
\end{ruledtabular}
\end{table}

\section{\label{sec:Methods}Methods}

\subsection{\label{sec:Analytical_Methods}Analytical methods}

\subsubsection{\label{sec:Finite-size_scaling}Finite-size scaling theory}

According to the finite-size scaling theory \cite{Finite-Size_Scaling_Theory_1990,Guide_to_Monte_Carlo_Simulations_2009} in the vicinity of the percolation threshold $p_c$ the probability of belonging to the largest cluster is
\begin{equation}
\label{eq:Pmax}
\mathcal{P}_{\max}(p;L)=\mathcal{S}_\text{max}(p;L)/L^5,
\end{equation}
where $\mathcal{S}_\text{max}$ is the size of the largest cluster, scales with the linear size of the system $L$ and the distance to the critical point $(p-p_c)$ as
\begin{equation}
\label{eq:scaling}
\mathcal{P}_{\max}(p;L) \cdot L^{\beta/\nu} = \mathcal{F}\left( (p-p_c)\cdot L^{1/\nu} \right),
\end{equation}
where $\beta$ is the critical exponent associated with the weight of the percolating cluster, the critical exponent $\nu$ describes the divergence of the correlation length and $\mathcal{F}$ is a scaling function (and usually unknown analytically).

At $p=p_c$ the rescaled probability of belonging to the largest cluster $\mathcal{P}_{\max}(p;L)\cdot L^{\beta/\nu}=\mathcal{F}(0)$ is independent of the size of the system $L$, which allows us to estimate the percolation threshold $p_c$ by searching the common point of $\mathcal{P}_{\max}(p;L)\cdot L^{\beta/\nu}$ versus $p$ plotted for various values of $L$.
An example of rescaled probability of belonging to the largest cluster $\mathcal{P}_\text{max}\cdot L^{\beta/\nu}$ as dependent on the site occupation probability $p$ for \textsc{sc(5)-1} lattice is presented in \Cref{fig:example_sc5-rescaledPmax_vs_p}.
The lattice names follow the convention proposed in Reference~\onlinecite{2010.02895} reflecting the lattice topology (here \textsc{sc(5)}, {\it i.e.}, simple cubic, five dimensions) and a numerical string specifying the coordination zones $i$, where the sites constituting the neighborhood come from (here 1).
The critical exponents in the scaling law \eqref{eq:scaling} for five-dimensional space were estimated as  
$\beta_5=0.8457$ and $\nu_5=0.5746$ \cite{PhysRevD.92.025012}, $\nu_5=0.5737(33)$ \cite{Zhang_2021}, $\nu_5=0.5720(43)$ \cite{Brzeski_2022}.

\subsubsection{\label{sec:Universality_of_pc}Universality of $p_c$ behavior}

The total coordination number of the lattice with predefined neighborhood is defined as 
\begin{equation} 
\label{eq:z} 
z=\sum_i z_i,
\end{equation}
where $z_i$ is number of sites in $i$-th coordination zone. 
One can also consider the weighted coordination number
\begin{equation} \label{eq:zeta} 
\zeta=\sum_i z_i r_i,
\end{equation}
where $r_i$ is the Euclidean distance from the sites in the $i$-th coordination zone to the central site \cite{2204.12593,2310.20668}.
Then, percolation thresholds follow \Cref{eq:pc_zeta-to-gamma} for two-dimensional lattices \cite{2006.15621}.

On the other hand, values $c=1.722$ and $b=1$ in \Cref{eq:pc_const-over-z} for \textsc{sc(5)} lattices with extended neighborhoods (up to the 7-th nearest neighbors) was reported in Reference \onlinecite{2308.15719}.

\subsubsection{\label{sec:Fractal_dimension_IPC}Fractal dimension $d_f$ of the incipient percolation cluster}

At percolation threshold $p=p_c$ the largest cluster of occupied sites (the incipient percolation cluster) has fractal properties \cite{Mandelbrot_1983,Fractals_for_the_Classroom_1,Fractals_for_the_Classroom_2} (see Reference~\onlinecite{Balankin_2024} for the most recent review). 
The size of the incipient percolation cluster scales with linear system size $L$ as 
\begin{equation}
\label{eq:M_vs_L}
\mathcal S_\text{max}\propto L^{d_f}.
\end{equation} 
For standard percolation the fractal dimension of incipient percolation cluster obeys scaling relation 
\begin{equation}
\label{eq:d_f_star}
    d_f^*=d-\beta_d/\nu_d,
\end{equation}
where $d$ is physical space dimension in which percolation takes place.
For two-dimensional space ($d=2$) we know exact values of $\beta_2=5/36$ and $\nu_2=4/3$ \cite[p.~52]{bookDS}, what yields exact value of fractal dimension $d_f^*(d=2)=91/48$.
The universality of value $d_f$ for two- and three-dimensional lattices and complex neighborhoods was recently confirmed in Reference \onlinecite{Krawczyk_2025}.
For five-dimensional space ($d=5$) and recent estimations of exponents $\beta_5=0.8457$ \cite{PhysRevD.92.025012} and $\nu_5=0.5720(43)$ \cite{Brzeski_2022} we find $d_f^*(d=5)\approx 3.5215$.

\subsection{\label{sec:Computational_Methods}Computational methods}


To calculate $\mathcal{S}_\text{max}$ we apply the \citeauthor{NewmanZiff2001} algorithm \cite{NewmanZiff2001}. 
\citeauthor{Krawczyk_2025} \cite{Krawczyk_2025} say: 
`The algorithm is fast and efficient as it is based on the recursive construction of the system with $n$ occupied sites with the addition of only one occupied site to the system containing $(n-1)$ already occupied sites.
Then convolution with the binomial (Bernoulli) probability distribution allows us to transform $\mathcal S_\text{max}(n)$ in the domain of the number $n$ of occupied sites into $\mathcal{S}_\text{max}(p)$ in the domain $p$ of the probability of site occupation.
The immanent part of the \citeauthor{NewmanZiff2001} algorithm \cite{NewmanZiff2001} is also a concept of the efficient construction of binomial coefficients.'

\begin{figure}
\includegraphics[width=.48\textwidth]{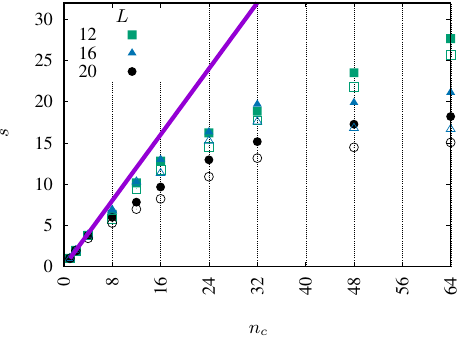}
\caption{\label{fig:speedup}
Parallel speedup $s$ for system sizes $L=12$, $16$, $20$ obtained from $R=192$ simulations per data point for $1\le n_c\le 64$ cores. Results for \text{sc(5)-1,2,3,4,5,6,7} are shown with filled symbols, while \textsc{sc(5)-1,7} is shown with open symbols. The solid line represents the ideal linear speedup
}
\end{figure}

The \citeauthor{NewmanZiff2001} algorithm originally implemented in C \cite{NewmanZiff2001}  was reimplemented in C++ and is available in a public repository~\cite{SM} for \textsc{sc(5)-1,2,3,4,5,6,7} as \texttt{sc-5-1234567.cpp}. An example showing the necessary modifications to obtain the source code for any other neighborhood is also given (for \textsc{sc(5)-1,7} as \texttt{sc-5-17.cpp}). 
To enable the large-scale computations required for complex neighborhoods, we utilized OpenMP parallelization \cite{Dagum_1998}.
The parallel speedup for two test cases, \textsc{sc(5)-1,7} and \textsc{sc(5)-1,2,3,4,5,6,7}, was measured across $R=192$ simulations performed for $L=12, 16$, and $20$.
The results of these tests, obtained on a 4.1~GHz Intel Xeon Platinum 8562Y+ 64-core CPU, are presented in \Cref{fig:speedup}.
While all curves are consistent with the trends predicted by Amdahl's law \cite{Amdahl_1967}, larger systems experience memory demands that scale rapidly with $L$.
This induces memory bandwidth bottlenecks and high-latency input/output operations (swapping/paging).

In total, the calculations presented in this work consumed $19$ million core-hours on the PLGrid Ares supercomputer \cite{ares}, which would have been equivalent to $\approx 2170$ years of single-core computation.

The choice of the OpenMP shared-memory approach was motivated by the necessity to execute a massive number of simulations ($R=10^6$).
To maximize hardware utilization and minimize computational wall-time, the calculations for each lattice type and size were distributed across up to 1,000 parallel batch jobs on the available infrastructure, depending on the memory requirements of the used lattice size.
This partitioning was also strictly required to manage memory constraints; for large lattice sizes, attempting to run a higher number of simulations concurrently on a single node would exceed the available RAM.
Each batch was handled by a single process that executed multiple simulations concurrently via OpenMP, utilizing all the available cores on a standard dual-socket compute node.
The data obtained from each batch were subsequently aggregated to yield the final results.

\section{\label{sec:Results}Results}

\subsection{Percolation threshold based on finite-size scaling theory}

With Newman--Ziff algorithm we compute size of the largest cluster $\mathcal{S}_\text{max}$, which allow for construction 
rescaled probability of belonging to the largest cluster $\mathcal{P}_\text{max}$ \eqref{eq:Pmax} as dependent on the site occupation probability $p$ (for example see \Cref{fig:example_sc5-rescaledPmax_vs_p}).
Repetitions of $R=10^5$ system construction give a clear common crossing point for various lattice sizes $L$ with $10^{-4}$ precision [see \Cref{fig:example_sc5-rescaledPmax_vs_p}(a)].
Ten times better statistics ($R = 10^6$) improve this accuracy to $10^{-5}$ [see \Cref{fig:example_sc5-rescaledPmax_vs_p}(b)] and we stick to this statistics as a compromise between expected accuracy and computation time.

The computed values of the percolation thresholds $p_c$ for various complex neighborhoods are presented in \Cref{tab:pc}.

\subsection{Universality of $p_c$ behavior}

\begin{figure}[tbp]
\includegraphics[width=.99\columnwidth]{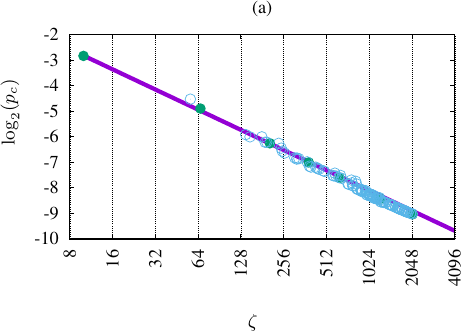}
\includegraphics[width=.99\columnwidth]{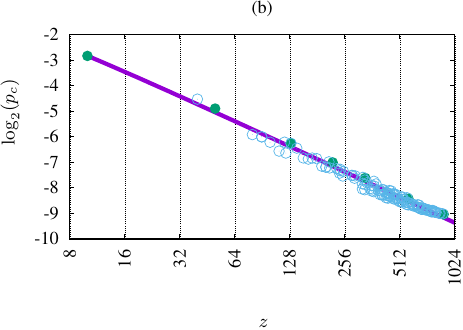}
\caption{\label{fig:zeta_vs_pc}Dependencies (a) $p_c(\zeta)=\zeta^{-g}$ (proposed in Reference~\onlinecite{2204.12593}) and (b) $p_c(z)=c/(z+b)$ (proposed in Reference~\onlinecite{2308.15719}). 
Open (full) symbols correspond to complex (extended) neighborhoods.
The least square method fits yield $g\approx 0.7913(43)$, $b\approx 1.01(18)$ and $c\approx 1.553(22)$}
\end{figure}

In \Cref{tab:pc} the computed values of the percolation thresholds $p_c$ are accompanied by the neighborhood characteristics $z$ \eqref{eq:z} and $\zeta$ \eqref{eq:zeta}.
This allows us to plot dependencies $p_c(\zeta)$ [\Cref{fig:zeta_vs_pc}(a)] and $p_c(z)$ [\Cref{fig:zeta_vs_pc}(b)].
Open (full) symbols correspond to complex (extended) neighborhoods.
The least squares method gives $g\approx 0.8293(24)$ in \Cref{eq:pc_zeta-to-gamma}, and $b\approx 1.01(18)$, $c\approx 1.553(22)$ in \Cref{eq:pc_const-over-z}.

\subsection{Fractal dimension $d_f$ of the incipient percolation cluster}

\begin{figure}[tbp]
\includegraphics[width=.99\columnwidth]{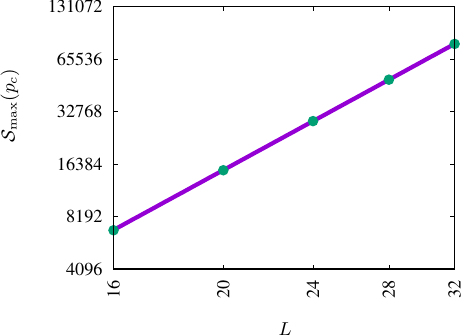}
\caption{\label{fig:df}Example of dependency of the size of the largest cluster $\mathcal{S}_\text{max}$ at percolation threshold $p_c$ on linear system size $16\le L\le 32$ for \textsc{sc(5)-1} neighborhood. 
The power law fit exponent gives system fractal dimension $d_f(\textsc{sc(5)-1})\approx 3.5453(30)$, which more or less agree with the earlier estimate $d_f(\textsc{sc(5)-1})\approx 3.5260(14)$ \cite{Zhang_2021} and theoretically predicted value $d_f^*=3.5215$ \eqref{eq:d_f_star}. 
The other dependencies $\mathcal{S}_\text{max}(p_c)$ vs. $L$ (for neighborhoods from \textsc{sc(5)-1,2} to \textsc{sc(5)-1,2,3,4,5,6,7}) are shown in 
Figure 6
in Supplemental Material \cite{SM}}
\end{figure}

To estimate the fractal dimension $d_f$ of the incipient percolation cluster we plot $\mathcal{S}_\text{max}$ at $p=p_c$ as dependent on the system linear size $L$.
An example of dependency of the size of the largest cluster $\mathcal{S}_\text{max}$ at percolation threshold $p=p_c$ on linear system size for $L=16$, 20, 24, 28, 32 for \textsc{sc(5)-1} is presented in \Cref{fig:df}.
The percolation thresholds $p_c$ are taken with ten times better accuracy than shown in \Cref{tab:pc} and based on values presented in References \onlinecite{PhysRevE.98.022120,2308.15719}.
The assumed values $p_c$ for which are calculated values of $d_f$ are collected in \Cref{tab:df}.
\Cref{tab:df} contains also the percentage error \cite[p. 14]{Abramowitz_1964} of fractal dimension $d_f$
\begin{equation}
\label{eq:delta}
\delta=\frac{d_f^*-d_f}{d_f^*} \cdot 100\%
\end{equation}
compared to the theoretically predicted \eqref{eq:d_f_star} value $d_f^*(d=5)\approx 3.5215$.

\begin{table}[tbp]
\caption{\label{tab:df}Estimated fractal dimension $d_f$ \eqref{eq:M_vs_L} and its percentage error $\delta$ \eqref{eq:delta} for various extended neighborhoods for \textsc{sc(5)} lattice. We assume theoretical value of $d_f^*(d=5)\approx 3.5215$ based on \Cref{eq:d_f_star} with $\beta_5=0.8457$ \cite{PhysRevD.92.025012} and $\nu_5=0.5720(43)$ \cite{Brzeski_2022}}
\begin{ruledtabular}
\begin{tabular}{rllll}
lattice \textsc{sc(5)-} & $p_c$ && $d_f$ & $\delta$\\ 
\hline 
1             & $0.14079633(4)$ &\cite{PhysRevE.98.022120} & $3.5453(30)$  & 0.68\% \\
1,2           & $0.033794(4)$   &\cite{2308.15719}         & $3.5307(23)$  & 0.26\% \\
1,2,3         & $0.013190(2)$   &\cite{2308.15719}         & $3.5525(16)$  & 0.88\% \\
1,2,3,4       & $0.007850(1)$   &\cite{2308.15719}         & $3.5579(11)$  & 1.03\% \\
1,2,3,4,5     & $0.0051575(5)$  &\cite{2308.15719}         & $3.5575(25)$  & 1.02\% \\
1,2,3,4,5,6   & $0.002981(1)$   &\cite{2308.15719}         & $3.57944(80)$ & 1.65\% \\
1,2,3,4,5,6,7 & $0.001934(1)$   &\cite{2308.15719}         & $3.5836(25)$  & 1.76\% \\
\end{tabular}
\end{ruledtabular}
\end{table}


\section{\label{sec:Discussion}Discussion}

The percolation thresholds for the neighborhoods on \textsc{sc(5)} lattice vary from $0.00193$ (for \textsc{sc(5)-1,2,3,4,5,6,7}) to $0.14079$  (for \textsc{sc(5)-1}). 
For extended (compact) neighborhoods, the obtained values $p_c$ agree within six digit accuracy with the percolation thresholds presented in Reference \onlinecite{2308.15719} as indicated in \Cref{tab:pc}. 

The dependencies $p_c(z)$ and $p_c(\zeta)$ can be nicely approximated by \Cref{eq:pc_const-over-z,eq:pc_zeta-to-gamma}, respectively.
The least-squares method leads to $g\approx 0.7913(43)$ and $c\approx 1.553(22)$ with the sum of the squares of the residuals equal to 
$8.1039\times 10^{-5}$ [when fitting the data to \Cref{eq:pc_zeta-to-gamma}] and $9.40327\times 10^{-5}$ [when fitting the data to \Cref{eq:pc_const-over-z}]. 
The present value $c\approx 1.553(22)$ in \Cref{eq:pc_const-over-z} improves the earlier estimate $c\approx 1.722$ obtained for extended (complex but compact) neighborhoods in Reference~\onlinecite{2308.15719}.

The fractal dimension calculated with the scaling relation $d_f^*(d=5)\approx 3.5215$ \eqref{eq:d_f_star} and does not differ much from the recently published $d_f=3.5260(14)$ in Reference \onlinecite{Zhang_2021}.
The percentage errors obtained for all the extended neighborhoods (complex and compact) considered here are $0.26\%\le\delta\le 1.76\%$.
The mean value of the computed fractal dimensions for seven extended neighborhoods is $\langle d_f\rangle=3.5581(70)$.

In summary, in this paper we show the results of massively parallel computations based on the Newman--Ziff algorithm implemented in C++ with OpenMP parallelization.
The obtained speed-up of computations reduces 2170 years of single-core computations---necessary for obtaining the results presented in this paper---much below typical half-decay time of the scientist.
127 thresholds for site percolation problem for complex neighborhoods in five-dimensional space are calculated (with 120 among them being estimated for the first time).
These percolation thresholds nicely follow the power law $p_c\propto\zeta^{-g}$ with $g\approx 0.7913(43)$.
This fit to heuristic prediction is qualitatively not worse than $p_c\propto(z+1)^{-1}$ (since both give the sum of the squares of the residuals of the order $\sim10^{-5}$).
For extended (complex and compact) neighborhoods, the mean value of the fractal dimension $\langle d_f\rangle=3.5581(70)$ calculated by definition \eqref{eq:M_vs_L} does not differ much from this predicted by the scaling relation \eqref{eq:d_f_star} $d_f^*(d=5)=3.5215$ (with percentage errors ranging from 0.26\% to 1.76\% for specific cases).

\begin{acknowledgments}
We gratefully acknowledge Polish high-performance computing infrastructure \href{https://ror.org/01m3qaz74}{PLGrid} (HPC Center: ACK Cyfronet AGH) for providing computer facilities and support within computational grant no. PLG/2026/019207.
This research was supported by a subsidy from the \href{https://ror.org/05dwvd537}{Polish Ministry of Science and Higher Education}.
\end{acknowledgments}

%

\end{document}